\documentclass[12pt,a4paper]{article}

\usepackage{epsfig}
\usepackage{amsmath}
\usepackage{verbatim}
\usepackage{psfrag}
\usepackage{bm}
\usepackage{dsfont}
\usepackage[footnotesize]{caption}

\newcommand{\dif}{\,\mbox{d}}

\newcommand{\oT}{T^*}
\newcommand{\MP}{M_{\text{P}}}

\newcommand{\ophi}{\phi^*}

\newcommand{\oS}{S^*}
\newcommand{\oN}{N^*}
\newcommand{\oW}{W^*}

\newcommand{\del}{\partial}

\newcommand{\euler}{\text{e}}

 \DeclareMathOperator{\diag}{diag}

\begin{document}

\begin{flushright}
\tt{CAFPE-103/08}\\
\tt{MPP-2008-108}\\
\tt{UGFT-233/08}
\end{flushright}

\vskip 0.75cm

\begin{center}

{\Large\bf Solving the $\eta$\,-\,Problem in Hybrid Inflation with Heisenberg Symmetry and
Stabilized Modulus}

\vskip 1cm

Stefan~Antusch,$^1$ Mar~Bastero-Gil,$^2$ Koushik~Dutta,$^1$ Steve~F.~King$^3$ \\ and Philipp~M.~Kostka$^1$\\[3mm]
{$^1$\it{
Max-Planck-Institut f\"ur Physik (Werner-Heisenberg-Institut),\\
F\"ohringer Ring 6,
80805 M\"unchen, Germany}\\[2mm]
$^2$\it{Departamento de Fisica Teorica y del Cosmos and
Centro Andaluz de Fisica de Particulas Elementales,
Universidad de Granada, 19071 Granada, Spain\\[2mm]
$^3$\it{School of Physics and Astronomy, University of Southampton,\\
Southampton, SO17 1BJ, United Kingdom}} }

\end{center}

\vskip 0.75cm

\begin{abstract}
We propose a class of models in which the $\eta$-problem of
supersymmetric hybrid inflation is resolved using a Heisenberg
symmetry, where the associated modulus field is stabilized and
made heavy with the help of the large vacuum energy during
inflation without any fine-tuning. The proposed class of models is
well motivated both from string theory considerations, since it
includes the commonly encountered case of no-scale supergravity
K\"ahler potential, and from the perspective of particle physics
since a natural candidate for the inflaton in this class of models
is the right-handed sneutrino which is massless during the
inflationary epoch, and subsequently acquires a large
mass at the end of inflation. We study a specific example motivated by  
sneutrino hybrid inflation with no-scale supergravity
in some detail, and show that the spectral index may lie within
the latest WMAP range, while the tensor-to-scalar ratio is very
small.
\end{abstract}

\date{\today}
\newpage

\section{Introduction}

The inflationary paradigm is extremely successful in solving the
horizon and flatness problems of the standard Big Bang cosmology,
and at the same time in explaining the origin of structure in the
observable
Universe~\cite{Liddle:2000cg}.
However the problem of how to incorporate inflation into a concrete model of high energy
particle physics remains unclear. On the observational side, the currently
available data is not precise enough to select a particular model,
whereas on the theoretical side we still lack the full
understanding of the dynamics of inflation. Among the several
mechanisms proposed for inflation, hybrid inflation \cite{Linde:1991km,
Copeland:1994vg} continues to be a very well motivated possibility
\cite{Linde:1997sj,Jeannerot:1997is, BasteroGil:1999fz},
especially from the point of view of constructing a model of
inflation based on a supersymmetric (SUSY)
extension of the Standard Model (SM)~\cite{Dvali:1994ms}.
The main advantage of hybrid inflation is that it involves small
field values below the Planck scale, thereby allowing a small
field expansion of the K\"ahler potential in the
effective supergravity (SUGRA) theory, facilitating the connection
with effective low energy particle physics models.
In this paper we will be concerned with two problems confronting
SUSY hybrid inflation,
namely the $\eta$-problem and the moduli stabilization problem,
and show that their joint resolution seems to favor
a particular class of models which are well motivated from both
string theory and particle physics.

To be consistent with recent observations \cite{Komatsu:2008hk,
Hinshaw: 2008 ap}, implementing inflation in a quantum field
theoretical context typically requires a scalar field (at least at
the effective field theory level), dubbed inflaton, whose
potential is extremely flat. Global supersymmetry provides a
plethora of additional scalar fields, i.e.\ the bosonic
superpartners of the SM fermions. However in order to have a flat
enough potential and be a viable candidate for being the inflaton,
either one (or several) of the scalar fields has to be a gauge
singlet or a flat direction in the multi-dimensional field space.
One promising candidate naturally emerges from the supersymmetric seesaw
mechanism, which provides a compelling and minimal explanation of
the smallness of the observed neutrino masses. The superfields
containing the SM-singlet right-handed neutrinos may be gauge singlets and their
bosonic components, the right-handed sneutrinos, may in principle play
the role of the inflaton in either chaotic inflation
\cite{Murayama:1992ua} or hybrid inflation \cite{Antusch: 2005 hp}.

In the framework of SUSY models of inflation it is necessary to take into
account the effects arising from the effective SUGRA theory,
which turn out to be important even for low field values. Such corrections
generically occur in locally supersymmetric (SUGRA) theory (including low energy effective
SUGRA theories arising from string theory) and threaten the flat direction for
inflation. The dangerous corrections arise from the effective SUGRA potential,
where the expansion of the K\"ahler potential generally leads to masses of the
order of the Hubble scale $H$ for all scalar fields,
including the inflaton. Such corrections to the inflaton mass would lead to a
slow-roll parameter $\eta \sim 1$ which would spoil inflation.
This is the so-called $\eta$-problem of SUGRA inflation \cite{Copeland:1994vg,
Dine:1995uk}.

Several attempts have been made to cure the $\eta$-problem as follows.
One option to treat the problem is to arrange for a small SUGRA
induced mass by hand, e.g.\ by a specific choice of the K\"ahler
potential or by a general expansion of the K\"ahler potential in
inverse powers of the Planck scale and finally tuning the
expansion parameters (see e.g.\ \cite{Murayama:1993xu,Antusch: 2005 hp} 
for examples in the context of sneutrino
inflation). However these choices are not motivated by any
symmetry argument. Another way to solve the $\eta$-problem is to
impose some symmetry requirement on the K\"ahler potential. For
example, shift symmetry has been used in several SUGRA inflation
model constructions \cite{Kawasaki:2000yn, Yamaguchi:2000vm,
Brax:2005jv}.
Another possibility is to use a Heisenberg symmetry\footnote{The
name Heisenberg symmetry is due to the invariance of the theory
under non-compact Heisenberg group transformations. An account is
given in e.g.~\cite{Binetruy: 1987}.} in the K\"ahler potential
which leads to a flat potential for the inflaton at
tree-level~\cite{Gaillard:1995az}. However, using a Heisenberg
symmetry in order to solve the $\eta$-problem requires the
introduction of a modulus field, which must be stabilized during
inflation.

The problem of combining moduli stabilization and inflation is
also common to compactifications of the higher dimensional
string theory. In the low energy effective 4-dimensional SUGRA
theory there are
several scalar fields, which are for instance related to the
geometry of the internal space of the higher dimensions, and which
are also called moduli fields. The requirements for the moduli
fields are exactly opposite to the ones for the inflaton field.
The moduli fields must be stable during inflation, in order not to
spoil inflation, and must remain stabilized after inflation. The necessity
of giving the moduli fields a large mass and stabilizing them at
comparatively large field values is called the ``moduli stabilization
problem" in the literature \cite{Brustein:1992nk, Dine:2000ds}.

It turns out that in potentially realistic SUGRA models it is hard
to stabilize the moduli and solve the $\eta$-problem of inflation
simultaneously. The moduli sector is usually not decoupled from
the inflaton sector, which means that any mechanism of moduli
stabilization would always have an effect on the inflaton sector.
In particular, a small variation of the moduli fields can give a
significant contribution to the inflationary potential, often
spoiling the conditions for inflation.
Discussions of the problems associated with moduli stabilization
and solving the $\eta$-problem, as well as directions for possible
solutions, can be found in the literature: For example, in
\cite{Ellis:2006ara} moduli stabilization is discussed in the
context of chaotic type potentials with the no-scale form of the
K\"ahler function. It has been noted that inflation may be
achieved if the associated moduli are fixed during inflation,
however at the expense of large fine-tuning. Similar conclusions
have been drawn recently in \cite{Davis:2008fv}. 
The possibility of modular inflation in this context has recently 
been discussed in~\cite{Covi:2008cn} and it has been pointed
out that subleading corrections to the no-scale K\"ahler potential 
could help to realize consistent inflation models. The effect of
couplings between moduli and inflaton sectors in hybrid models
has been discussed in \cite{Brax:2006ay}. There it has been found
that when the usual $\eta$-problem in SUGRA is solved by using a
shift symmetry, the moduli dynamics in general contributes a large
mass to the inflaton and spoils inflation. A possible resolution
to this problem has been proposed in \cite{Davis:2008sa}, again
using a shift symmetry. In the context of hybrid type of
inflationary models, the possibility of using a Heisenberg
symmetry for solving the $\eta$-problem and issues connected to
the associated moduli problem were discussed in
\cite{Gaillard:1998xx}, however no explicit model has been
considered.

In this paper we present a class of SUSY hybrid inflation
models in which the $\eta$-problem is solved by a
Heisenberg symmetry of the K\"ahler potential.
The associated modulus field is stabilized and made
heavy with the help of the large vacuum energy during inflation
without any fine-tuning.\footnote{
Here we do not address the problem of stabilizing the modulus 
after inflation, which we assume to be achieved by a different mechanism.} 
Because of the Heisenberg symmetry of the
K\"ahler potential, the tree-level potential of the inflaton is
flat and only lifted by radiative corrections, induced by
Heisenberg symmetry breaking superpotential couplings.
The resulting class of models are well motivated from the
point of view of string theory since they include the
case of no-scale SUGRA K\"ahler potentials which are
ubiquitous in string constructions.
The models are also
well motivated from the point of view of particle physics
since they allow the possibility that the inflaton may be
identified with the right-handed sneutrino in
SUSY see-saw models of neutrino masses, for example as
in the model of sneutrino hybrid inflation in
\cite{Antusch: 2005 hp}. We emphasize that the general class of models
considered here applies to a wider class of singlet inflaton models
with Heisenberg symmetry, where the inflaton field is
massless during the inflationary
epoch, and subsequently acquires a large mass at the end
of inflation. However much of the paper is devoted to
a particular example inspired by sneutrino hybrid inflation with no-scale
SUGRA, and which we discuss in some detail in order
to illustrate the approach. In the considered example we
demonstrate explicitly how the modulus gets
stabilized by the large vacuum energy density provided during
inflation, and find that the spectral index $n_s$ is predicted to be
below $1$, but above about $0.98$, while the tensor-to-scalar ratio
$r$ is below ${\cal O}(0.01)$. In extensions of this minimal model
we find examples where a spectral index as low as $0.95$ can be
realized.

The paper is organized as follows: In section \ref{The Model} the
general framework is outlined and our explicit example scenario is
presented. In section \ref{Diagonal  Basis} we describe the
background evolution of the fields for the considered
example. Section \ref{Tree-Level Scalar
Potential} contains the analysis of the relevant tree-level
potential including the relevant inflaton-dependent masses of the
scalar fields. The flatness of the tree-level potential is
lifted by the radiative corrections which are calculated in
section \ref{One-Loop Effective Potential}. Moreover, it is
devoted to numerical solutions of the inflationary dynamics,
including the stabilization of the modulus. The predictions of the
model are presented in section \ref{Observations}. Our Summary and
Conclusions are given in section \ref{Conclusions}.

\section{Framework}\label{The Model}
In the following,  $N$ will denote the chiral superfield which
contains the inflaton as scalar component. Furthermore, we will
introduce the superfields $H$ and $S$, where $H$ contains the
waterfall field of hybrid inflation and where the $F$-term of $S$
will provide the vacuum energy during inflation. In particle
physics applications $N$ may be identified with the
right-handed sneutrino, $H$ with a Higgs field which breaks
some high energy symmetry, and $S$ with some driving field
whose $F$ term drives the $H$ vacuum expectation
value (VEV).
However for simplicity,
we will consider SM singlet superfields throughout the paper,
noting that our discussion may be generalized to the case where
this assumption is dropped. We will consistently use the same
notation for the scalar component of the superfield and the
superfield itself.

\subsection{General Class of Models}\label{GeneralScenario}
We start by considering the following
general framework where the superpotential has the form
\begin{eqnarray}\label{Eq:W_general}
W= \kappa\,S\left(g_1(H,N)-M^2\right)+ g_2(H,N)\,,
\end{eqnarray}
and where the K\"ahler potential is of the type\footnote{Here and throughout
the paper we use units where the reduced Planck mass 
$\MP \approx 2.4 \times 10^{18}\,\text{GeV}$ is set to one.}
\begin{eqnarray}\label{Eq:Kaehler_general}
K= (|S|^2 + |H|^2  +  \kappa_{S}\, |S|^4 +  \kappa_{SH}\, |S|^2
|H|^2 + \dots) + g_3(\rho) |S|^2 + f(\rho)\,.
\end{eqnarray}
$\rho$ contains the inflaton $N$ as well as a modulus field $T$ in
a combination which is invariant under Heisenberg symmetry in
order to solve the $\eta$-problem for $N$ and is defined as
\begin{equation}\label{Heisenberg}
\rho=T+\oT-|N|^2\,.
\end{equation}
For explicitness, we have written the part of the K\"ahler potential
in Eq.~(\ref{Eq:Kaehler_general}) which contains only the fields
$H$ and $S$ as an expansion in powers of $\MP^{-1}$. $g_1$, $g_2$
are functions of $H$ and $N$ and $g_3$ and $f$ are functions of
$\rho$. $\kappa$, $\kappa_{S}$ and $\kappa_{SH}$ are dimensionless
parameters.

In the scenario we have in mind the scalar component of $S$ only
contributes the large vacuum energy during inflation by its
F-term, but remains at zero during inflation. The waterfall field
$H$ is responsible for ending inflation by a second order phase
transition when it develops a tachyonic instability at some critical
value of $N$. Below this critical value, $H$ acquires a large VEV
determined by the scale $M$ and also gives a large
mass to the inflaton $N$. As main features of the general
framework we require that, as in sneutrino hybrid inflation,
$W=0$, $W_N =W_H=W_T=0$, $W_S \not=0$ but $H=S=0$ during
inflation. It has been emphasized in \cite{Stewart: 1995 hp} that
these criteria are desirable for solving the $\eta$-problem using
a Heisenberg symmetry.

Before we discuss an explicit example where we demonstrate how the
modulus is stabilized consistent with inflation, let us discuss in
general terms the requirements on the functions $g_1$, $g_2$,
$g_3$ and $f$. To start with, $g_1$ has to be chosen such that the
vacuum energy during inflation is provided by the $F$-term
$|F_S|^2 = |g_1(H=0,N)-M^2|^2$. Typically, $g_1$ depends only on
$H$, and we note that it may also contain effective couplings like
$H^4/\Lambda^2$ (as e.g.\ in \cite{Antusch: 2005 hp}). $g_2$ has
to lead to a  positive $N$-dependent mass squared for $H$ via
$|F_H|^2 + |F_N|^2$ during inflation. Examples for possible $g_2$
are terms like $N^m H^2$ with $m \ge 1$. If, on the other hand,
only one power of $H$ appears in a term of $g_2$, this would give
a tree-level contribution to the $N$-potential which may spoil
inflation. Finally, $g_3$ together with $f$ shape the potential
for $\rho$ (which contains the modulus $T$ and which, as we will
show, is the field which has to be stabilized during inflation in
order to solve the moduli problem in the context of inflation).
The main idea here is that the K\"ahler potential term $g_3(\rho)
|S|^2$ will induce a contribution to the potential of the order of
the vacuum energy $\sim |F_S|^2$ during inflation and can (in
combination with a suitably chosen $f(\rho)$, e.g.\ of no-scale
form) efficiently stabilize $\rho$ during inflation. The
stabilization of the modulus during inflation is thus driven by a
different mechanism than after inflation and in the following we
will assume that additional terms in the superpotential or
K\"ahler potential which stabilize the modulus after inflation can
be neglected during inflation.

\subsection{Explicit Example Inspired by Sneutrino Inflation}\label{Explicit Example}
The explicit example model which we will investigate in
the remainder of the paper
is defined by the superpotential\footnote{The example is inspired
by the model of sneutrino hybrid inflation in 
\cite{Antusch: 2005 hp} where $N$ is the singlet sneutrino superfield, however we
emphasize that it can be any SM-singlet superfield. To be precise,
the considered superpotential is a variation of that used in the
model of \cite{Antusch: 2005 hp} where instead of
$H^2$ in the bracket a more complicated term $\frac{H^4}{M'{}^2}$ was considered.
We would like to note
that we have verified that a variant of the model, where the
coupling $\frac{\lambda}{M_{*}}N^2 H^2$ is replaced by the
renormalizable coupling $\lambda N H^2$, leads to similar results.} 
\begin{equation}\label{Superpotential}
W=\kappa\,S\left(H^2-M^2\right)+\frac{\lambda}{M_{*}}N^2H^2\,,
\end{equation}
and the K\"ahler potential
\begin{equation}\label{Kaehlerpotential}
K(H,S,N,T)\equiv |H|^2 + \left(1+
\kappa_{S}\,|S|^2+\kappa_{\rho}\,\rho\,\right) |S|^2 + f(\rho)\,,
\end{equation}
where $\kappa$, $\lambda$, $\kappa_{S}$ and $\kappa_{\rho}$ are
dimensionless parameters and $M_{*}$ is a mass scale. This
particular form of the superpotential of
Eq.~\eqref{Superpotential} and K\"ahler potential of
Eq.~\eqref{Kaehlerpotential} can be obtained with $\kappa_{SH} =
0$, $g_1 = H^2$,  $g_2 = \frac{\lambda}{M_{*}}N^2H^2$, and $g_3 =
\kappa_{\rho}\,\rho $ from the general framework of
Eqs.~\eqref{Eq:W_general} and \eqref{Eq:Kaehler_general} in the
last subsection.  The first term of Eq.~\eqref{Superpotential} is
the standard SUSY hybrid inflation term, with the difference that
$S$ stays at zero both during and after inflation while $H$ is
kept at zero during inflation but acquires a VEV when $N$ drops
below a critical value. This term essentially provides a large
vacuum energy density during inflation and a VEV for $H$ after
inflation. The second term induces a mass for $H$ during inflation
when $N\ne0$, which keeps it at zero. After inflation, the VEV
$\langle H\rangle=\mathcal{O}(M)$ gives a mass to $N$. The purpose
of the coupling $\kappa_{S}$ is to give a large mass for the
$S$-field, which keeps it at zero both during and after inflation.
We have not included the term proportional to $\kappa_{SH}$, since
it is optional in the sense that it is not required for the model
to work. For $\kappa_{SH} = \mathcal{O}(1)$ or below, the term has
negligible effect on the predictions for the observable
quantities. However, $\kappa_{SH} \approx \mathcal{O}(10)$ allows
to lower the predictions for the spectral index, as will be
discussed in section~\ref{Observations}.

Finally, the additional coupling constant $\kappa_{\rho}$ which
admits a coupling between the combined modulus $\rho$ and $S$ is
needed in order to generate the stabilizing minimum for $\rho$.
After transforming to the basis where $\rho$ and $N$ are the
independent degrees of freedom (DOFs) the potential for $N$ is
flat at tree-level due to the Heisenberg symmetry.

\section{Background Evolution}\label{Diagonal Basis}
In this section, we derive the background equations of motion (EOMs) of
all the relevant fields and calculate the tree level potential. We
also describe how to transform to the $(N,\rho)$-basis and why
this basis is convenient. We assume that $S=H=0$
during inflation, such that $W=0$ and all derivatives
$W_{H}=W_{N}=W_{T}=0$ except for $W_{S}\ne 0$. In
section~\ref{Tree-Level Scalar Potential}, we will explicitly show
from the full scalar potential that the aforementioned assumptions
are justified.

Working in a flat Friedmann-Lema\^itre-Robertson-Walker
Universe with a metric\\
$g_{\mu\nu}=\diag{\left(1,-a^2(t),-a^2(t),-a^2(t)\right)}$ and
minimal coupling to gravity, the $N=1$ SUGRA action is given by
\begin{equation}\label{Action}
S\left[\Phi_{i}\right]=\int \text{d}^4x\,\sqrt{-g}\,
\mathcal{L}_{\text{SUGRA}}
\left(\phi_{i},\del_{\mu}\,\phi_{i},\chi_{i},\del_{\mu}\,\chi_{i}\right)\,,
\end{equation}
where $g=\det{\left(g_{\mu\nu}\right)}=-a^6(t)$ and
$\mathcal{L}_{\text{SUGRA}}$ is the SUGRA invariant Lagrangian
density with the bosonic component fields $\phi_{i}$ and the Weyl
spinor fermionic superpartners $\chi_{i}$ of the chiral
superfields $\Phi_{i}$. $t$ denotes cosmic time.

The fermion mass terms, which we will need in the calculation of
the loop-corrections in section~\ref{One-Loop Effective
Potential}, are given by \cite{ Brax:2006ay, Ferrara: 1994 ht}
\begin{equation}\label{fermionmassterm}
\mathcal{L}_{\text{SUGRA}}\supset -\,\frac{1}{2}\,m_{3/2}
\left(G_{ij}+G_{i}G_{j}-G_{ij\bar{k}}G^{\bar{k}}\right)
\chi_{i}\,\chi_{j}-\text{H.c.}\,.
\end{equation}
Here, the K\"ahler function $G$ and the gravitino mass $m_{3/2}$
are defined as
\begin{equation}
G=K+\ln|W|^2\,,\qquad m_{3/2}=|W|\,\text{e}^{K/2}\,.
\end{equation}
This leads to a fermionic mass matrix written in terms of
derivatives of the superpotential and K\"ahler potential as
\begin{equation}\label{fermionmassmatrix}
\left(\mathcal{M}_{\text{F}}\right)_{ij}=
\euler^{K/2}\left(W_{ij}+K_{ij}\,W+K_{i}\,W_{j}
+K_{j}\,W_{i}+K_{i}\,K_{j}\,W
-{K^{k\bar{l}}}\,K_{ij\bar{l}}\,\mathcal{D}_{k}W\right)\,.
\end{equation}

The scalar part of the Lagrangian density in a $N=1$ SUGRA theory
reads
\begin{equation}\label{SUGRALagrangian}
\mathcal{L}_{\text{SUGRA}}\supset
\mathcal{L}_{\text{Kin}}-V_{F}\,,
\end{equation}
with scalar kinetic terms and F-term scalar potential,
respectively, given by
\begin{equation}\label{KineticTerms}
\begin{split}
\mathcal{L}_{\text{Kin}}&=g^{\mu\nu}\,K_{i\bar{j}}\,
\left(\del_{\mu}\,\phi_{i}\right)\left(\del_{\nu}\,\ophi_{j}\right)\,,\\
V_{\text{F}}&=\text{e}^K\left[K^{i\bar{j}}\,
\mathcal{D}_{i}W\,\mathcal{D}_{\bar{j}}\oW
 - 3|W|^2\right]\,.
\end{split}
\end{equation}
Considering that all the chiral superfields are gauge singlets,
D-term contributions to the potential are absent. Here, indices
$i,j$ denote the different scalar fields and lower indices on the
superpotential or K\"ahler potential represent the derivatives
with respect to the associated chiral superfields or their
conjugate where a bar is involved. The inverse K\"ahler metric is
dubbed $K^{i\bar{j}}=K^{-1}_{i\bar{j}}$. Also, in
Eqs.~\eqref{KineticTerms} and~\eqref{fermionmassmatrix} we have
used the definition
\begin{equation}
\mathcal{D}_{i}W:=W_{i}+K_{i}\,W\,.
\end{equation}

The K\"ahler metric can be calculated as the second derivative of
the K\"ahler potential in Eq.~\eqref{Kaehlerpotential} with
respect to the superfields and their conjugates which in
$(H,S,N,T)$-basis reads
\begin{equation}\label{Kaehlermetric}
 \left(K_{i\bar{j}}\right)=
 \begin{pmatrix}
 1 & 0 & 0 & 0\\
 0 & 1+\kappa_{\rho}\,\rho+4\,\kappa_{S}\,|S|^2 & -\kappa_{\rho}\,N\,\oS & \kappa_{\rho}\,\oS\\
 0 & -\kappa_{\rho}\,\oN\,S & f''(\rho)|N|^2 - f'(\rho) - \kappa_{\rho}\,|S|^2
 & -f''(\rho)\oN \\
 0 & \kappa_{\rho}\,S & -f''(\rho)\,N & f''(\rho)
 \end{pmatrix}\,.
 \end{equation}
With $S=H=0$ during inflation, this reduces to the block-diagonal
form
 \begin{equation}\label{KaehlermetricInflation}
 \left(K_{i\bar{j}}\right)=
 \begin{pmatrix}
 1 & 0 & 0 & 0\\
 0 & 1+\kappa_{\rho}\,\rho & 0 & 0\\
 0 & 0 & f''(\rho)|N|^2 - f'(\rho) & -f''(\rho)\oN \\
 0 & 0 & -f''(\rho)\,N & f''(\rho)
 \end{pmatrix}\,,
 \end{equation}
which suggests that the $(N,T)$-sub-block can be treated
independently. Since $S$ basically remains static during and after
inflation, we do not take its EOM into
account. The kinetic sector of the waterfall field $H$ decouples
from $(N,T)$ and its kinetic term is canonical.

Since the phases of the scalar fields $S,\,H$ and $N$ as well as $\text{Im}(T)$
very quickly approach a constant value
in an expanding Universe and subsequently decouple from the 
absolute values and $\text{Re}(T)$ in the EOMs
(as will be discussed in detail in the Appendix), 
we only consider the absolute values and $\text{Re}(T)$ in
what follows and denote them by lowercase letters
$s=\sqrt{2}\,|S|$,
$h=\sqrt{2}\,|H|$,
$n=\sqrt{2}\,|N|$ and
$t/2=\sqrt{2}\,\text{Re}(T)$.
The phases and $\text{Im}(T)$, we set constant
(or without loss of generality to zero). 
In addition, the spatial
derivatives satisfy $\mathbf{\nabla}\phi_{i}=0$ in a homogeneous,
isotropic Universe. The kinetic terms for $t$ and $n$ are then
obtained to be
\begin{equation}\label{KinetictermsReal}
\mathcal{L}_{\text{kin}}=\frac{f''(\rho)}{4}\,
n^2\left(\del_{\mu}n\right)^2
-\frac{f'(\rho)}{2}\left(\del_{\mu}n\right)^2
-\frac{f''(\rho)}{2\sqrt{2}}\,n\, \del_{\mu}n\,\del^{\mu}t
+\frac{f''(\rho)}{8}\left(\del_{\mu}t\right)^2\,.
\end{equation}
In order to transform to the independent DOFs $\rho$ and $n$, we
use the definition in Eq.~\eqref{Heisenberg} and end up with the
kinetic Lagrangian terms
\begin{equation}\label{KinetictermsDiagonal}
\mathcal{L}_{\text{kin}}=\frac{f''(\rho)}{4}\left(\del_{\mu}\rho\right)^2
-\frac{f'(\rho)}{2}\left(\del_{\mu}n\right)^2\,,
\end{equation}
which are diagonal in the field derivatives $\del_{\mu}\rho$ and $\del_{\mu} n$.

Upon variation of the action given in Eq.~\eqref{Action} and
introduction of the Hubble scale $H(t)=\dot{a}(t)/a(t)$, we obtain
the EOMs \footnote{Note from the EOMs
in Eq.~\eqref{EOMs} that for $f'(\rho_{0})=0$ there is a
divergence in the acceleration of $n$. This can be avoided for
non-vanishing $\kappa_{\rho}$, such that the minimum of the
potential $\rho_{\text{min}}$ is shifted away from the minimum of
$f(\rho)$ and thus $\rho_{\text{min}}\ne\rho_{0}$. As we will see
in the next section, $f(\rho)$ does not even have to have a
minimum in order to stabilize $\rho$.} for the classical scalar
fields
\begin{equation}\label{EOMs}
\begin{split}
\ddot{n}+3H(t)\dot{n}+\frac{f''(\rho)}{f'(\rho)}\,\dot{\rho}\,\dot{n}-\frac{1}{f'(\rho)}\frac{\del V}{\del n}&=0\,,\\
\ddot{\rho}+3H(t)\dot{\rho}+\frac{f^{(3)}(\rho)}{2f''(\rho)}\,\dot{\rho}^2
+\dot{n}^2+\frac{2}{f''(\rho)}\frac{\del V}{\del\rho}&=0\,.
\end{split}
\end{equation}
For the simulation of the evolution of the scale factor during
inflation, we add the Friedmann equation
\begin{equation}\label{friedmann}
\dot{a}(t)=a(t)H(t)\,,\qquad H(t)=\sqrt{\frac{\varepsilon}{3}}\,,
\end{equation}
where the energy density in terms of non-canonically normalized
fields is given by\\ $\varepsilon=\mathcal{L_{\text{kin}}}+V$. In
our case, the potential is only determined by the F-terms
$V=V_{\text{F}}$.

With $S=H=0$ during inflation, the tree-level F-term scalar
potential in Eq.~\eqref{KineticTerms} reduces to the simple form
\begin{equation}\label{FtermpotentialInflation}
V_{\text{tree}}=V_{\text{F}}=\text{e}^{f(\rho)}\, K^{-1}_{S\oS}\,
\left|\frac{\del W}{\del S}\right|^2 =\kappa^2M^4\cdot
\frac{\text{e}^{f(\rho)}} {\left(1+\kappa_{\rho}\,\rho\right)}\,.
\end{equation}
From Eq.~(\ref{KinetictermsDiagonal}) we can see that in order to 
have a positive kinetic term for the inflaton field in
the potential minimum, the function $f(\rho)$ should fulfill the
requirement that $f'(\rho_{\text{min}})$ is negative.

As mentioned in the beginning of this chapter, we want to
summarize why the new $(n,\rho)$-basis is more convenient. We have
shown in this section that in this basis, the K\"ahler metric
diagonalizes during inflation, which is a great simplification. In
addition, the tree-level potential is exactly flat in
$n$-direction. This is due to the Heisenberg symmetry which
protects $n$ from obtaining large mass corrections in the SUGRA
expansion. Thus, the $\eta$-problem of SUGRA inflation has a
simple solution. Moreover, as we will see in the next section,
besides the K\"ahler metric, the mass matrices are simultaneously
diagonal in this basis. Owing to the diagonal K\"ahler metric, the
kinetic energy is diagonal in the $(n,\rho)$-basis and thus the
standard formalism of calculating the effective potential from
radiative corrections applies \footnote{Apart from the
normalization factor.}, which is well known in the
literature~\cite{Coleman: 1973}. Hence, the one-loop radiative
corrections are easy to calculate. Having everything diagonalized
in the independent fields $(n,\rho)$, we consider this basis to be
the physically relevant one. It is important to note that even
though the kinetic energies of the fields are diagonal, they are
still not yet canonically normalized except for the field $h$. We
will transform to the normalized fields for $\rho = \rho_{min}$ in
the next section.

\section{Analysis of the Tree-Level Scalar Potential with No-Scale Supergravity}\label{Tree-Level Scalar Potential}
As mentioned before, this section is dedicated to the classical
tree-level F-term scalar potential. The assumptions $S=H=0$ and
thus $W=W_{\phi_{i}}=0$ for all $\phi_{i}\ne S$ used in
section~\ref{Diagonal Basis} must be proven from the full scalar
potential. This is justified, if the potential has minima in all
relevant directions at $s=h=0$ with masses of the fields larger
than the Hubble scale $m^2_{\phi_i}> H^2$.

Using the SuperCosmology code~\cite{Kallosh: 2004 ht} we
calculate the F-term scalar potential from
Eq.~\eqref{KineticTerms}, and since the potential is very lengthy
and not illuminating, we will not explicitly write down the
result. But we will show all results derived from the potential.

Before proceeding further we need to specify the function
$f(\rho)$. One such example of the function that we mostly work
with is the following no-scale form
\begin{equation}\label{specificf(rho)}
f(\rho)=-\,3\,\ln{(\rho)}.
\end{equation}
We emphasize that this is only one specific choice within the
class of models where the K\"ahler potential has the form of
Eq.~\eqref{Eq:Kaehler_general}. Making the curvature of the
potential along the $\rho$-direction larger than the Hubble scale
renders the modulus stable very quickly. Now, the term
proportional to $\kappa_{\rho}$ generates the minimum of the
potential by switching on the coupling between $S$ and $\rho$.

First, we have checked that both $s$ and $h$ have a minimum at
$s=h=0$ due to
\begin{equation}
\frac{\del V}{\del s}\Big{|}_{s=h=0}= \frac{\del V}{\del
h}\Big{|}_{s=h=0}=0\,.
\end{equation}
After transforming the potential to the $(n,\rho)$-basis by the
substitution $t\rightarrow\rho+n^2/2$, the curvature of the
potential along the modulus field direction around $s=h=0$ is
given by
\begin{equation}\label{rhomass}
\frac{\del^2V}{\del\rho^2}\Big{|}_{s=h=0}=
\frac{2\,\kappa^2M^4\,\text{e}^{f(\rho)}}{\rho^2\left(1+\kappa_{\rho}\rho\right)^3}
\left[\,6+15\,\kappa_{\rho}\,\rho
+10\,\kappa_{\rho}^2\,\rho^2\right]\,.
\end{equation}
The field $s$ is also supposed to stay at its minimum during and
after inflation. For the curvature along the $s$ direction, we
obtain
\begin{equation}\label{smass}
\frac{\del^2 V}{\del s^2}\Big{|}_{s=h=0}=
\frac{\kappa^2M^4\,\text{e}^{f(\rho)}}
{3\left(1+\kappa_{\rho}\rho\right)^2}
\left[\,-12\,\kappa_{S}+\left(3+4\kappa_{\rho}\rho\right)^2\right]\,.
\end{equation}
Finally, the waterfall field $h$ has the curvature
\begin{equation}\label{hmass}
\frac{\del^2V}{\del h^2}\Big{|}_{s=h=0}=
\text{e}^{f(\rho)}\left[\frac{\lambda^2}{M^2_{*}}\,n^4+
\frac{2\,\left(\kappa
M\right)^2}{\left(1+\kappa_{\rho}\rho\right)}
\left(\frac{M^2}{2}-1\right)\right]\,.
\end{equation}
Strictly speaking, these values of the curvatures cannot be
interpreted as the squared masses $m_{\phi_i}^2$ of the respective
fields as the fields are not yet canonically normalized, except
for the waterfall field $h$. From
Eq.~\eqref{KinetictermsDiagonal}, we know that the normalization
depends on the $\rho\,$-modulus only, and as we will see soon, it
settles to its minimum at the very beginning of inflation. We will
justify it both by comparing the mass of the $\rho\,$-modulus at
the minimum to the Hubble scale, and also by looking at the full
evolution of the fields by solving Eqs.~\eqref{EOMs}. After the
$\rho\,$-modulus has settled to its minimum we can easily
canonically normalize the fields and this normalization typically
makes changes of $\mathcal {O}(1)$. Note that only the curvature
of the $h$ field depends on the field value of the inflaton $n$.
Therefore it will be the only considerable contribution to the
one-loop effective potential. We can also easily verify that all
the cross terms vanish. Therefore the full mass matrix is diagonal
\begin{equation}
\mathcal{M}^2\big{|}_{s=h=0} =
\diag{\left(m^2_{\tilde{h}}\,,\,m^2_{\tilde{s}}\,,\,0\,,\,m^2_{\tilde{\rho}}\right)}\,,
\end{equation}
with a completely flat $n$-direction (where $m^2_{\tilde{n}}=0$),
as expected. From now on we denote all canonically normalized
fields with a tilde.

\begin{figure}
\psfrag{rho}{$\rho\,[\MP]$} \psfrag{pot}{$V(\rho)\,[\kappa^2M^4]$}
\center
\includegraphics[scale=0.9]{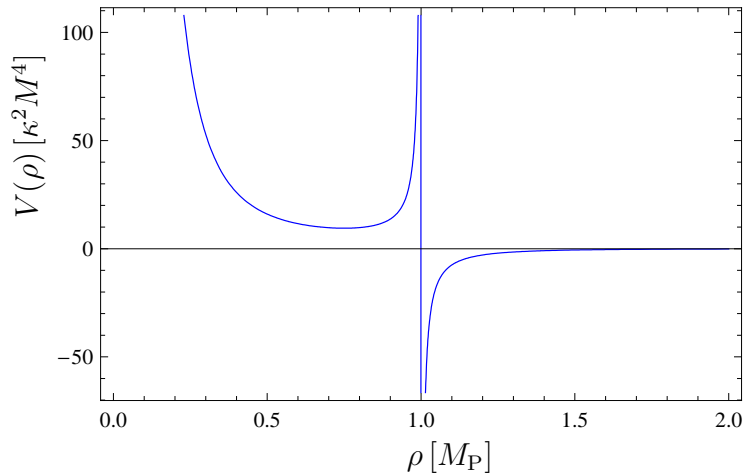}
\caption{\label{modpot}Tree-level scalar potential depending on
$\rho$ for $\kappa_{\rho}=-1$. $\rho$ is given in units
of the reduced Planck mass.}
\end{figure}

Depending on the choice of $\kappa_{\rho}$ and hence the minimum
$\rho_{\text{min}}$, the other masses can be fairly large in the
inflationary trajectory. The potential at $s=h=0$
is therefore given by Eq.~\eqref{FtermpotentialInflation} together
with the no scale form of $f(\rho)$ in Eq.~\eqref{specificf(rho)}
and depicted in Fig.~\ref{modpot}.
As we can see from Eq.~\eqref{FtermpotentialInflation}, for the modulus
to be stable during inflation, the initial field value of $\rho$ must be less
than $-\kappa^{-1}_{\rho}$.
The potential then gets minimized at
\begin{equation}\label{rhominimum}
\rho_{\text{min}}=-\frac{3}{4\,\kappa_{\rho}}\,.
\end{equation}
At the minimum, the canonically normalized fields in terms of the non-canonically
normalized ones are given by the following relations:
$\tilde{s} = \frac{s}{2}$, $\tilde{\rho} = \sqrt{8/3} \,\rho$,
$\tilde{n} = 2\,n $, and $\tilde h = h $. In this stable patch,
the masses of the scalars at the minimum in
Eqs.~\eqref{rhomass},\eqref{smass} and \eqref{hmass} reduce to
\begin{equation}\label{specificmasses}
\begin{split}
m^2_{\tilde \rho}&=-\frac{16384}{81}\,\kappa_{\rho}^{5}\,\kappa^2M^4\,,\\
m^2_{\tilde s}&=\frac{4096}{27}\,\kappa_{\rho}^{3}\,\kappa_{S}\,\kappa^2M^4\,,\\
m^2_{\tilde
h}&=\,-\frac{64}{27}\,\kappa_{\rho}^{3}\,\left[\frac{\lambda^2}{16\,M^2_{*}}
\,\tilde n^4+ 8\,\kappa^2 M^2\left(\frac{M^2}{2}-1\right)\right]\,,\\
m^2_{\tilde n}&=0\,.
\end{split}
\end{equation}
To see that these are stable during inflation, we need to compare
them to the squared Hubble scale in the same patch, given by
\begin{equation}
H^2=\frac{1}{3}V\left(\rho_{\text{min}}\right)\big{|}_{s=h=0}
=-\frac{256}{81}\,\kappa_{\rho}^{3}\,\kappa^2M^4\,\,.
\end{equation}
For the squared modulus mass, the requirement $m^2_{\tilde
\rho}/H^2>1$ is easily fulfilled, since $m^2_{\tilde
\rho}/H^2=64\,\kappa_{\rho}^{2}$ and the condition is thus
satisfied if $|\kappa_{\rho}|>1/8$. Since only the case of a
negative sign generates a minimum in the potential, we can even
require $\kappa_{\rho}<-1/8$. The $\tilde s$ field can be heavier
than the Hubble scale if the condition $\kappa_{S}<-1/48$ holds.

In this model, the waterfall mechanism works in the usual way.
From Eq.~\eqref{specificmasses}, it is clear that the mass of the
waterfall field can be arbitrarily high if the field value of $
\tilde n$ is large enough. Once $ \tilde n$ drops below its
critical value $\tilde n_{\text{c}}$ at which $m^2_{\tilde h}=0$,
the waterfall field gets destabilized and slow-roll inflation
ends. From Eq.~\eqref{specificmasses} the critical value of the
waterfall field can be found to be
\begin{equation}
\tilde{n}^2_{\text{c}}=8\,\frac{\kappa}{\lambda}\left(MM_{*}\right)
\sqrt{2-M^2}\,.
\end{equation}

\section{One-Loop Effective Potential}\label{One-Loop Effective Potential}
Having shown that all fields are stabilized during inflation in
the inflationary trajectory $s=h=0$ and that the inflaton
direction $n$ is exactly flat at the classical level, we calculate
the one-loop radiative corrections to the effective potential in
this section.
These corrections are induced by Heisenberg symmetry
breaking superpotential couplings in combination with broken SUSY
during inflation, and will serve to generate a slope for the inflaton field.

It is important to generate such a tilted potential for two reasons.
Firstly in order to let the inflaton field start
rolling in the first place, such that it reaches its critical
value at some point in time and inflation can end. Secondly,
an exactly flat potential would produce primordial density
perturbations with a scale-invariant spectrum and thus a spectral
index $n_{\text{s}}=1$. Such a flat spectrum would contradict current
observations in the CMB photons as observed by WMAP~\cite{Komatsu:2008hk,Hinshaw:
2008 ap} and is excluded at the $95$\% confidence
level (CL).

The Coleman-Weinberg one-loop radiative correction to the
effective potential in a supersymmetric theory~\cite{Coleman:
1973, Weinberg:1973ua, Gamberini:1989jw} is given by
\begin{equation}
V_{\text{loop}}(\tilde{n})=\,\frac{1}{32\pi^2}\,\text{Str}
\left[\mathcal{M}^2(\tilde n)\,Q^2\right]
\,+\,\frac{1}{64\pi^2}\,\text{Str} \left[\mathcal{M}^4(\tilde n)
\left(\ln{\left(\frac{\mathcal{M}^2(\tilde n)}{Q^2}\right)}
-\frac{3}{2}\right)\right]\,,
\end{equation}
where $\mathcal {M}$ is the mass matrix and $Q$ is the
renormalization scale. It is important to note that we are
evaluating the effective potential in the approximation that the
$\rho$ field has stabilized to its minimum. For
$\rho=\rho_{\text{min}}$, only $h$ contributes
$\tilde{n}$-dependent mass terms to the effective potential.

Upon introduction of the new dimensionless variable
\begin{equation}
x:={\left(\frac{\lambda}{\kappa}\right)}^2
\frac{1+\kappa_{\rho}\,\rho}{2\,\left(MM_{*}\right)^2}
\,n^4\,,
\end{equation}
the squared masses are of a simple form.
The bosonic contribution comes from the scalar and pseudoscalar
masses of the $h$ field, which from Eq.~\eqref{hmass} are given by
\begin{equation}
m^2_{\text{B}}=2\,\frac{\left(\kappa M\right)^2}
{\left(1+\kappa_{\rho}\,\rho\right)}\,\text{e}^{f(\rho)}
\left[x+\frac{M^2}{2}\mp1\right]\,,
\end{equation}
where the minus refers to the scalars and the plus to the
pseudoscalars. In the considered case, the mass of the fermionic
superpartner from Eq.~\eqref{fermionmassmatrix} reduces to
$m_{\text{F}}=\text{e}^{K/2}\,W_{HH}$. Hence, the squared fermion mass is
obtained to be
\begin{equation}
m^2_{\text{F}}=2\,\frac{\left(\kappa M\right)^2}
{\left(1+\kappa_{\rho}\,\rho\right)}\,\text{e}^{f(\rho)}\,x\,.
\end{equation}

Taking into account the spin-multiplicity for the fermions, the
resulting one-loop correction is given by
\begin{equation}\label{Vloop}
\begin{split}
V_{\text{loop}}(x)&= \frac{\left(\kappa M\right)^4}
{64\left(1+\kappa_{\rho}\,\rho\right)^2\pi^2}\,
\Big[\,\\
&\,4\,\text{e}^{2f(\rho)}\,\left(x+M^2/2-1\right)^2
\Big[\ln{\left(\frac{2\,\kappa^2 M^2\,\text{e}^{f(\rho)}}
{\left(1+\kappa_{\rho}\,\rho\right)\,Q^2}\right)}
+\ln{\left(x+M^2/2-1\right)}-3/2\Big]\\
+&\,4\,\text{e}^{2f(\rho)}\,\left(x+M^2/2+1\right)^2
\Big[\ln{\left(\frac{2\,\kappa^2 M^2\,\text{e}^{f(\rho)}}
{\left(1+\kappa_{\rho}\,\rho\right)\,Q^2}\right)}
+\ln{\left(x+M^2/2+1\right)}-3/2\Big]\\
-&\,8\,\text{e}^{2f(\rho)}\,x^2\,\Big[\ln{\left(\frac{2\,\kappa^2
M^2\,\text{e}^{f(\rho)}}
{\left(1+\kappa_{\rho}\,\rho\right)\,Q^2}\right)}+
\ln{\left(x\right)}-3/2\Big]\Big]\\
+&\,\frac{\left(\kappa M Q\right)^2}
{16\left(1+\kappa_{\rho}\,\rho\right)\pi^2}
\Big[\text{e}^{f(\rho)}M^2\Big]\,.
\end{split}
\end{equation}

We now would like to make a few clarifying remarks concerning the
calculation of the one-loop effective potential.

First of all, neglecting all mass eigenvalues besides the ones for
$H$ is justified, since under the assumption that $\rho$ has
settled to its VEV, all other terms are independent of $n$ and
therefore just contribute a constant energy density which adds to
$V_{\text{tree}}$. Fixing the renormalization scale
$Q=m_{\text{F}}/\sqrt{x}$ as we do for the predictions in
section~\ref{Observations}, it turns out that all these
contributions can be safely neglected w.r.t. the tree-level
potential given in Eq.~\eqref{FtermpotentialInflation}. This is
even true for the $Q^2$-term in Eq.~\eqref{Vloop}.

Furthermore, we are aware of the fact that there is a remaining
$Q$-dependence in the observables. However, using sensible values
of $Q$ around the scale of inflation, a change of $Q$ only results
in a shift of the model parameters (due to the renormalization
group flow). The predictions for the observable quantities do not
change by a noteworthy amount.

Moreover, as all the observables are calculated at horizon exit,
i.e around 50 to 60 e-folds before the end of inflation, for all
practical purposes we substitute $\rho = \rho_{\text{min}}$ in the
above expression to find the observables. Strictly speaking, to
calculate the one-loop potential for a dynamical $\rho$, one would
need to keep both $n$ and $\rho$ canonically normalized at every
moment in time.

In order to show that the assumption $\rho=\rho_{\text{min}}$ is a
legitimate one, we numerically simulate the full evolution of the
non-canonically normalized fields from the EOMs of
Eq.~\eqref{EOMs} using
\begin{equation}\label{effectivePot}
V_{\text{eff}}(n,\rho)=V_{\text{tree}}(\rho)+V_{\text{loop}}(n,\rho)\,,
\end{equation}
with rather general initial field values.

\begin{figure}
\psfrag{MP}[][bl][1][90]{\hspace{0.0cm}$\rho,n\:[\MP]$} \psfrag{rho}{$\rho$}
\psfrag{n}{$n$} \psfrag{Ne}{\hspace{-1.9cm}$H\, t\,$ ($\simeq
N_{\text{e}}$ for $H\,t\ge 5$)} \center
\includegraphics[scale=0.9]{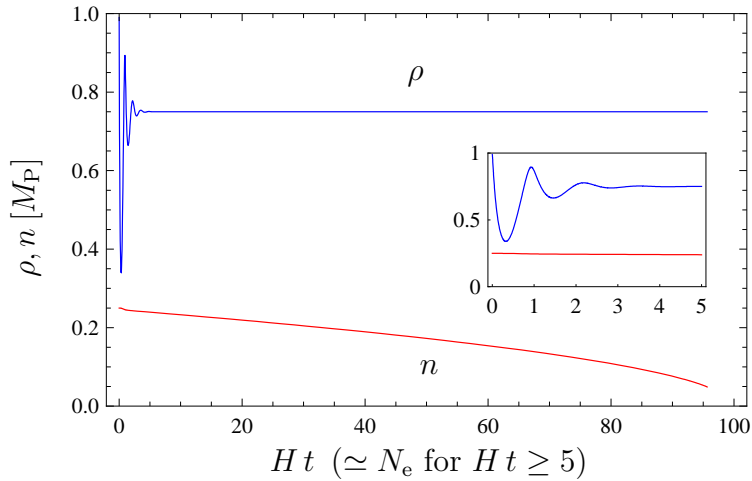}
\caption{\label{evol}Evolution of the modulus field $\rho$ (blue)
and the inflaton field $n$ (red) as a function of $H\,t$. The
inlay shows the behavior of the fields for the period $H\,t\le 5$
during which $\rho$ settles to its minimum. $\rho$ and $n$ are given in units
of the reduced Planck mass.}
\end{figure}

One example of such a numerical solution is shown in
Fig.~\ref{evol}. We can see that $\rho$ indeed settles to its
minimum very quickly and we can achieve a large enough number of
e-folds of inflation with $n$ moving to smaller values while
$\rho$ is stabilized at the minimum of its potential. For
producing the plot we have chosen example model parameters $\kappa
=0.05$ and $\lambda / M_* = 0.2$, which are compatible with the
observational constraints as will be discussed in the next
section.

\section{Predictions}\label{Observations}
In order to explain the observations, a viable model of inflation
has to account for the spectral index $n_{\text{s}}$ and the
amplitude of the curvature perturbations $P^{1/2}_{\mathcal{R}}$
as observed by WMAP~\cite{Komatsu:2008hk,Hinshaw: 2008 ap}. In
addition we can calculate the tensor-to-scalar ratio $r$ and the
running of the spectral index
$\text{d}n_{\text{s}}/\text{d}\ln{k}$. Each of these quantities
has to be evaluated at horizon exit, i.e.\ about $50$ e-folds
before the end of slow-roll inflation.

The slow-roll parameters are defined as
\begin{equation}
\epsilon=\frac{1}{2}\left(\frac{V'}{V}\right)^2\,,\quad
\eta=\left(\frac{V''}{V}\right)\,,\quad
\xi^2=\left(\frac{V'V'''}{V^2}\right)\,.
\end{equation}
With these, the observables are given by
\begin{equation}\label{observables}
\begin{split}
n_{\text{s}}&\simeq 1-6\,\epsilon+2\,\eta\,,\\
r&\simeq16\,\epsilon\,,\\
\frac{\text{d}n_{\text{s}}}{\text{d}\ln{k}}&\simeq
16\,\epsilon\,\eta-24\,\epsilon^2-2\,\xi^2\,.
\end{split}
\end{equation}
In addition, the amplitude of the curvature perturbations can be
obtained from
\begin{equation}\label{power}
P^{1/2}_{\mathcal{R}}=\frac{1}{2\sqrt{3}\,\pi}\,
\frac{V^{3/2}}{|V'|}\,,
\end{equation}
where in our case $V=V_{\text{tree}}+V_{\text{loop}}$. At the
$68$\% CL, the spectral index is measured to be
$n_{\text{s}}=0.960^{+0.014}_{-0.013}$ and the amplitude of the
spectrum $P^{1/2}_{\mathcal{R}}\simeq(5.0\pm0.1)\cdot 10^{-5}$,
while the evidence for a running of the spectral index remains
very weak ($\dif n_{\text{s}}/\dif\ln{k}=-0.032^{+0.021}_{-0.020}$). The new limit
on the tensor-to-scalar ratio is $r<0.2$ at the $95$\% CL
\footnote{These values are found using combined data from WMAP,
Type Ia supernovae and Baryon Acoustic Oscillations~\cite{Komatsu:2008hk,Hinshaw: 2008 ap}.}.

In order to obtain the predictions of the considered model, we
calculate the observables from the full loop-corrected potential.
We want to stress the fact that all fields besides the inflaton 
direction $n$ acquire a constant value very quickly such that the model
can effectively be treated as a single-field model of inflation.
Hence, Eqs.~\eqref{observables},~\eqref{power} apply and there is no
curving of the trajectory in field space and thus no isocurvature mode.
Therefore we fix $\rho=3/4$ to its minimum for $\kappa_{\rho}=-1$ (c.f.\ Eq.~(\ref{rhominimum})).
Since only the combination $\lambda/M_{*}$ is relevant, we can fix
$M_{*}=1$ without loss of generality. For each point in parameter
space, the scale $M$ is numerically calculated at horizon exit
such that the amplitude of the curvature perturbations
$P^{1/2}_{\mathcal{R}}$ resembles the observed value to one sigma.
In addition, the renormalization scale is taken to be
$Q=m_{\text{F}}/\sqrt{x}$ which makes the constant
log-contribution vanish in the loop-potential Eq.~\eqref{Vloop}.

\begin{figure}
\psfrag{nc}{\hspace{0.06cm}$\tilde{n}_{\text{c}}$}
\psfrag{ne}{\hspace{0.06cm}$\tilde{n}_{\text{e}}$}
\psfrag{pot}{$V\,[10^{-12}\,\MP^4]$}
\psfrag{Vloop}{\hspace{-0.8cm}$V_{\text{tree}}+V_{\text{loop}}$}
\psfrag{Vtree}{$V_{\text{tree}}$}
\psfrag{nnorm}{$\tilde{n}\,[\MP]$} \center
\includegraphics[scale=0.9]{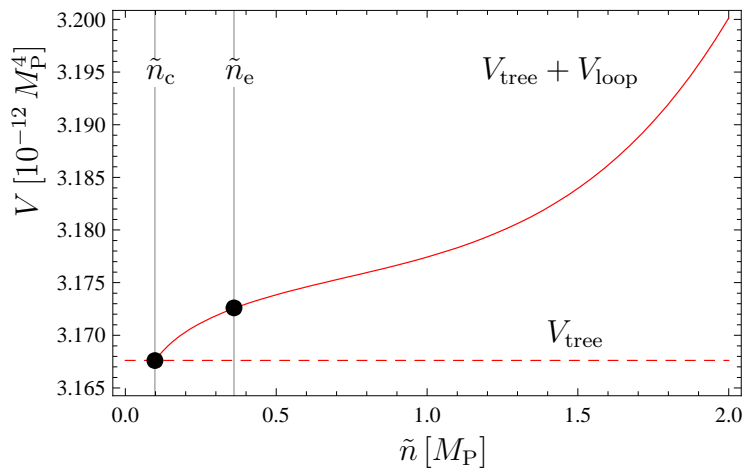}
\caption{\label{looppot}Graphical illustration of the one-loop
effective potential for $\tilde{n}$ with typical values of the
field 50 e-folds before the end of inflation
$\tilde{n}_{\text{e}}$ and at the critical value
$\tilde{n}_{\text{c}}$ where inflation ends. $\tilde{n}$ is given in units
of the reduced Planck mass.}
\end{figure}

As an example, we take a point in the remaining two-dimensional
parameter space. It is given by $\left(\kappa,\lambda\right)=
\left(0.05,\,0.2\right)$. The dependence of the effective
loop-corrected potential on the canonically normalized inflaton
$\tilde{n}$ is depicted in Fig.~\ref{looppot}. We integrated the
slow-roll EOMs in order to obtain the field value $50$ e-folds
before the field reaches the critical value
$\tilde{n}_{\text{c}}\simeq 0.10$, which is given by
$\tilde{n}_{50}\simeq 0.36$. As can be seen from the potential
form, inflation occurs well below the inflection point located
around $\tilde{n}=1$. The curvature and hence $\eta$ is negative
in this region and $\epsilon \ll |\eta|$, which implies that the
spectral index $n_s$ is below $1$.

From Eqs.~\eqref{observables} and~\eqref{power}, we obtain for the
spectral index and the scale $M$ from the COBE normalization
evaluated $50$ e-folds before the end of inflation
\begin{equation}
n_{\text{s}}\simeq 0.982\,,\qquad \frac{M}{\MP}\simeq 3.4\cdot
10^{-3}\,.
\end{equation}
This is not within the $68$\% CL of the WMAP 5-year
data, but still within 95\% CL~\cite{Hinshaw: 2008 ap}.
In addition, the tensor-to-scalar ratio and the running of the
spectral index are obtained to be
\begin{equation}
r\simeq 9.0\cdot 10^{-5}\,,\qquad
\frac{\text{d}n_{\text{s}}}{\text{d}\ln{k}}\simeq -2.4\cdot
10^{-3}\,,
\end{equation}
which are both rather small.
As typical for an effective single-field inflation
model, the non-Gaussianity parameter $f_{\text{NL}}$
is negligible.

\begin{figure}[!h]
\psfrag{r}{\large$r$} \psfrag{ns}{\large$n_{\text{s}}$}
\psfrag{lambda}{\hspace{0.2cm}$\lambda$} 
\psfrag{kappa}{\hspace{0.2cm}$\kappa$}
\psfrag{M}{$M/\MP$} \center
\includegraphics[scale=0.5]{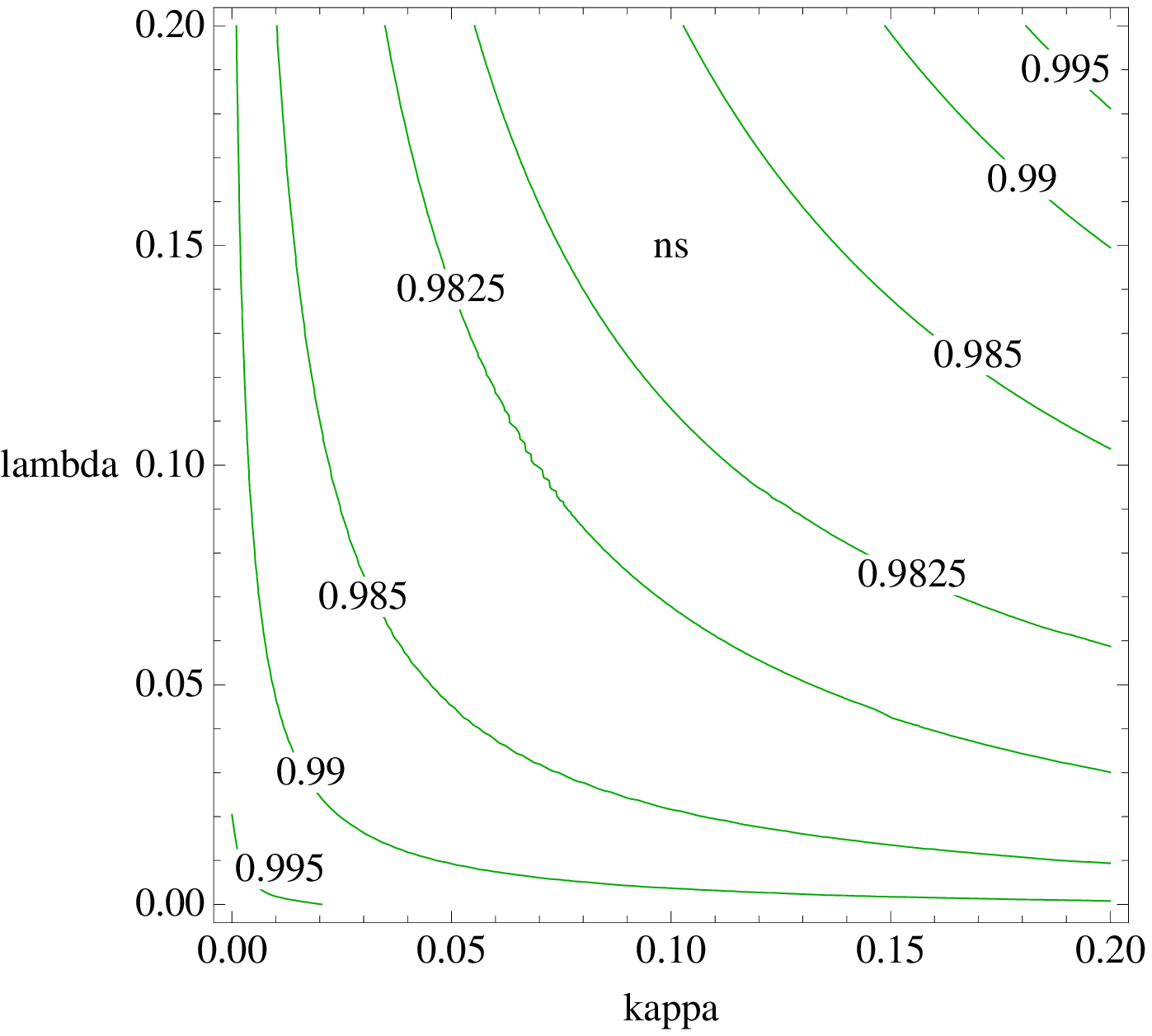}
\includegraphics[scale=0.5]{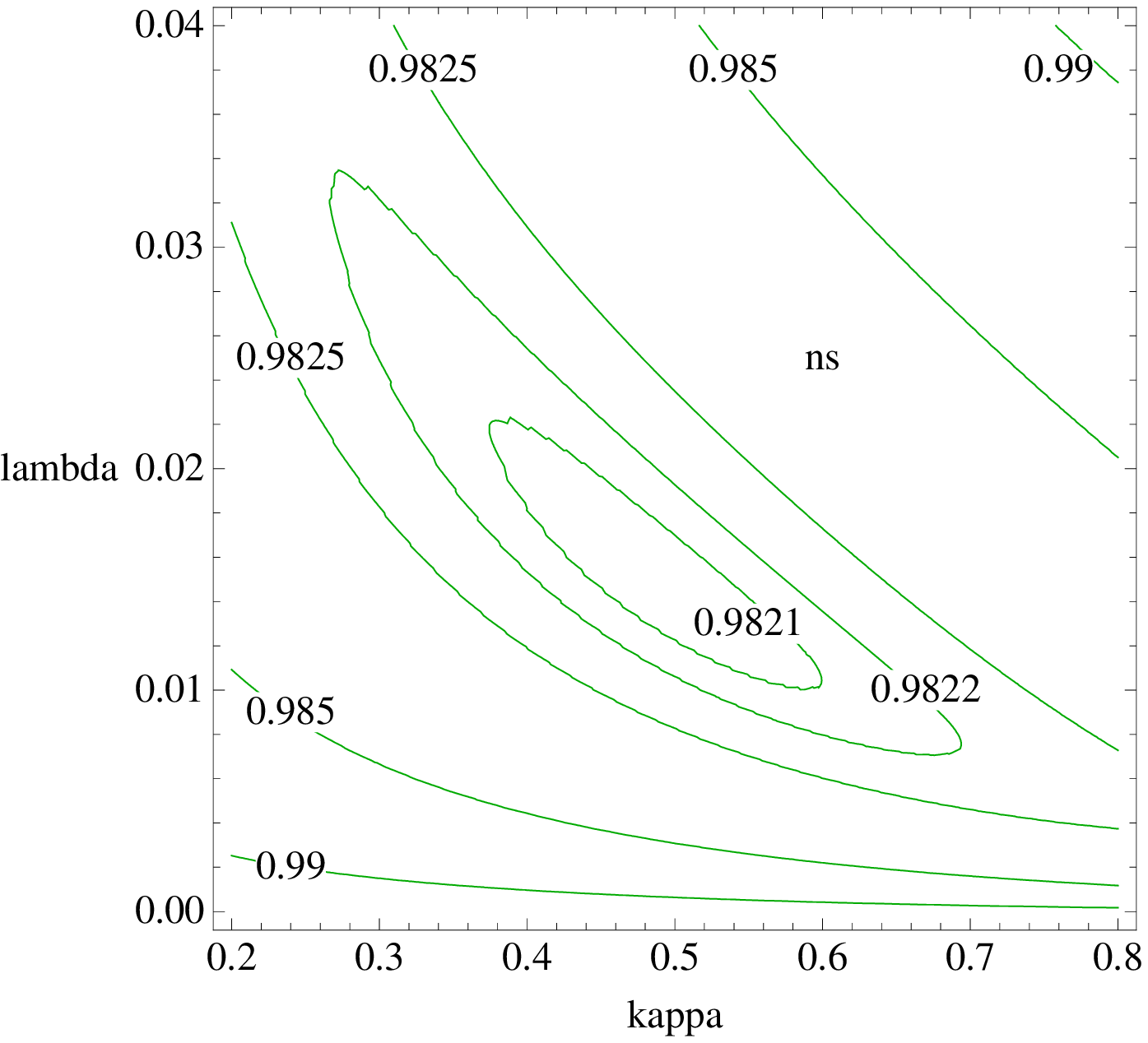}\\
\includegraphics[scale=0.5]{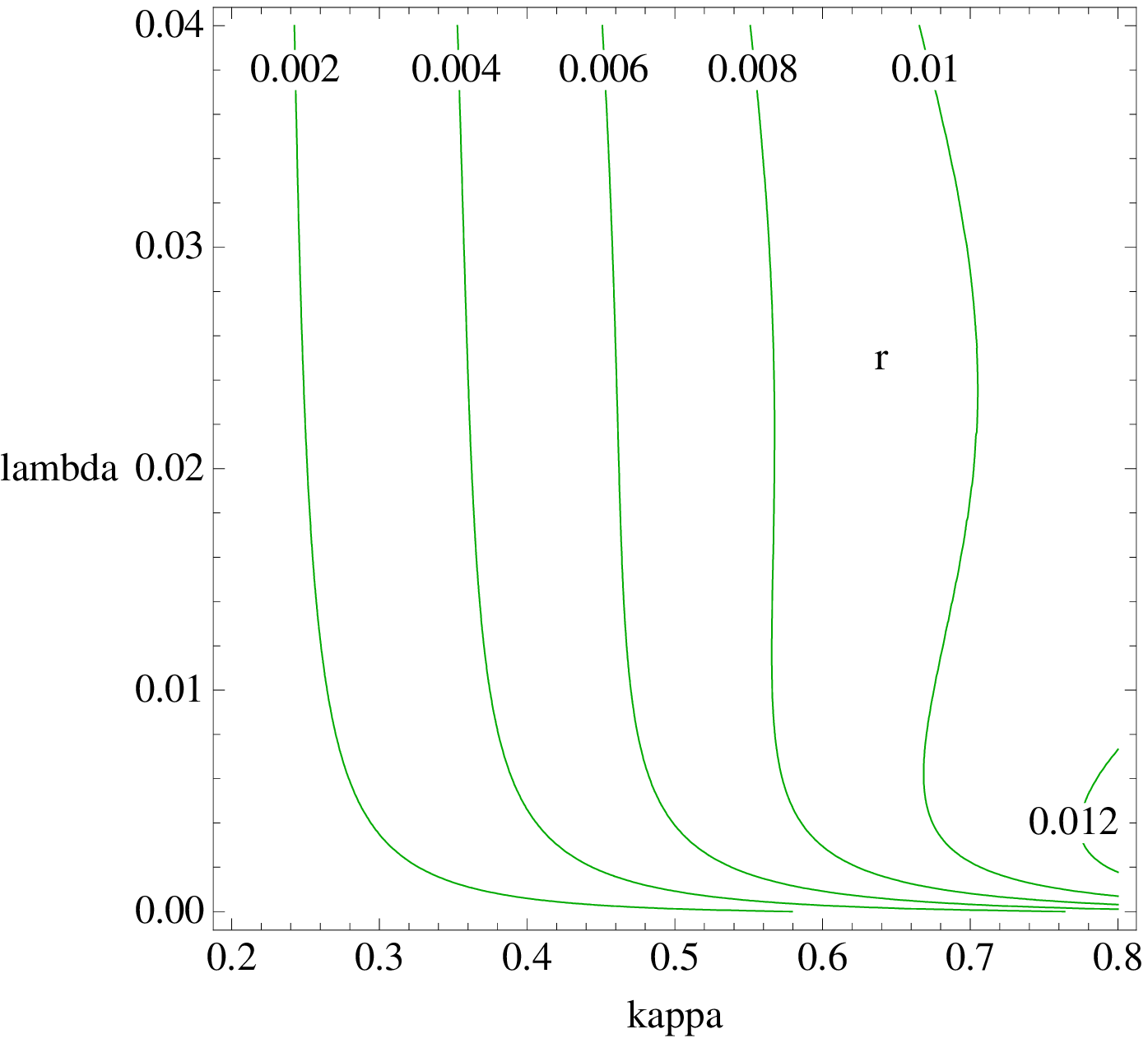}
\includegraphics[scale=0.5]{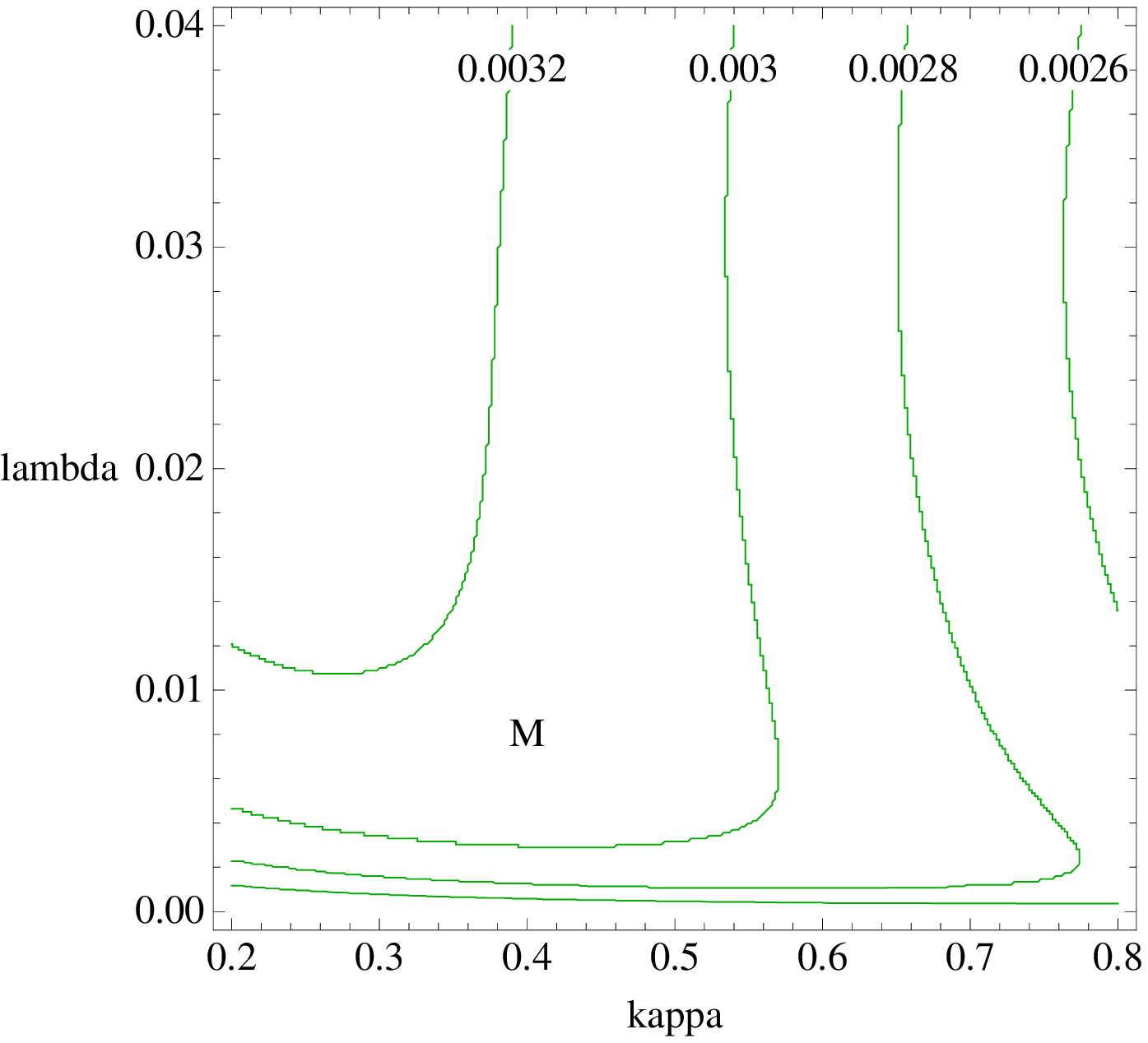}
\caption{\label{paramspace}Contours of the predicted values of
$n_{\text{s}}$, $r$ and $M$ depending on $\kappa$ and $\lambda$.}
\end{figure}

In order to investigate the parameter space, and give the
predictions for the spectral index and the tensor-to-scalar ratio
in this model, we scan this two-dimensional space. Therefore, we
again fix the other parameters and the renormalization scale as
above. The results are displayed in Fig.~\ref{paramspace}. In the
upper left corner, the contour lines of the spectral index are
plotted over a wider range of the parameter space, where both
$\lambda$ and $\kappa$ have been varied from $0$ to $0.2$. The
other three plots show contour lines of $n_{\text{s}}$, $r$ and
the scale $M/\MP$ in the intervals in which $\lambda$ has been
varied from $0$ to $0.04$ and $\kappa$ from $0.2$ to $0.8$, where
a minimum of the spectral index has been found. 
In the shown ranges, 
the spectral index $n_s$ is found to be below $1$, but
above about $0.98$. The tensor-to-scalar ratio $r$ is below ${\cal
O}(0.01)$, and $M/\MP$ is ${\cal O}(10^{-3})$.

We would like to stress that the above results have been
calculated using the minimal model defined in
Eqs.~\eqref{Superpotential} and~\eqref{Kaehlerpotential} with
$f(\rho)$ being of no-scale form as given in Eq.~(\ref{specificf(rho)}). 
Although the no-scale form is particularly well motivated, in general 
this assumption might be relaxed and a different function $f(\rho)$
might be chosen. The main requirement for $f(\rho)$ is that the 
potential for $\rho$ has a minimum of the order of the Planck scale 
and that the shape of the potential 
forces $\rho$ to settle rapidly at its minimum. After $\rho$ has settled
at its minimum, the values of $f(\rho_{\text{min}})$ and its derivatives 
affect the normalization of the inflaton field and also the field-dependent 
masses which finally enter the loop potential. 
We have analyzed some examples with generalized functions
$f(\rho)$ and found that in the considered cases the shape of the 
potential was not affected and the effects on the observables were 
small. To give one explicit example, for $f(\rho)=1/\rho$ and $\kappa_{\rho}=-1$,
we find a minimum at $\rho_{\text{min}}=(\sqrt{5}-1)/2$ where the modulus
stabilizes quickly such that inflation can occur.
The minimal value for the spectral index is again around $n_{\text{s}}\sim 0.98$
and the tensor-to-scalar ratio as well as the scale of inflation and the
running of the spectral index are only slightly changed.
However, a full exploration of general functional dependences 
of $f(\rho)$ is beyond the scope of the paper. 

On the contrary, as noted already in section~\ref{Explicit Example}, 
we find that the inclusion of additional couplings, for instance of
$\kappa_{SH}\ne 0$ as in Eq.~\eqref{Eq:Kaehler_general}, could
lower the spectral index at loop-level. The reason is that with such
modifications, the form of the potential changes qualitatively
compared to Fig.~\ref{looppot}.
For example, for the parameters 
$(\kappa,\lambda,\kappa_{SH})=(0.05,0.2,10)$ we find a
spectral index of $n_{\text{s}}\simeq0.953$ at horizon exit where
the COBE normalization fixes $M/\MP\simeq 0.0029$.

\section{Summary and Conclusions}\label{Conclusions}
In this paper we have proposed a class of models in which the
$\eta$-problem of SUSY hybrid inflation
is resolved using a Heisenberg symmetry, where the
associated modulus field is stabilized and made
heavy with the help of the large vacuum energy during inflation
without any fine-tuning.
The proposed class of models is well motivated both from string theory
considerations, since it includes the commonly encountered case of
no-scale SUGRA K\"ahler potential, and
from the perspective of particle physics since
a natural candidate for the inflaton in this class of models is the
right-handed sneutrino, i.e.\ the superpartner of the SM-singlet right-handed
neutrino, which is massless during the inflationary
epoch and subsequently acquires a large mass at the end
of inflation.

In order to illustrate the approach we have developed in
some detail a specific example motivated by sneutrino hybrid inflation
with no-scale SUGRA. In this example the right-handed sneutrino field
$N$ appears in the Heisenberg combination in
Eq.~\eqref{Heisenberg}, the
superpotential has the form in
Eq.~\eqref{Superpotential}, and the K\"ahler potential has the
form in Eq.~\eqref{Kaehlerpotential} where in practice we have
assumed the no-scale form in Eq.~\eqref{specificf(rho)}. In this
model the singlet $S$ is stabilized during inflation due to its
non-canonical K\"ahler potential, and since the right-handed
sneutrino contained in $N$ has its mass protected by the Heisenberg symmetry,
it becomes a natural candidate for the inflaton, with the
associated modulus field stabilized and made heavy with the help
of the large vacuum energy during inflation (where we have not
addressed the problem of the stability of the modulus after
inflation). Because of the Heisenberg symmetry the tree-level
potential of the sneutrino inflaton is flat and only lifted by
radiative corrections (induced by Heisenberg symmetry breaking
superpotential couplings) which we have studied and found to play
a key part in the inflationary dynamics.

We have found that in the considered setup the spectral index
$n_s$ typically lies below $1$ with typical values shown in
Fig.~\ref{paramspace} for the case of the model defined in
Eqs.~\eqref{Superpotential} and~\eqref{Kaehlerpotential} with
$f(\rho)$ given in Eq.~(\ref{specificf(rho)}). However,
the inclusion of additional couplings, for instance of
$\kappa_{SH}\ne 0$ as in Eq.~\eqref{Eq:Kaehler_general}, could
lower the spectral index at the loop-level. For example, for the
parameters $(\kappa,\lambda,\kappa_{SH})=(0.05,0.2,10)$ we have found a
spectral index of $n_{\text{s}}\simeq 0.953$ and $M$ fixed to $M/\MP\simeq 0.0029$
by the COBE normalization.
We expect further changes to the
predictions for the observables in other extensions or variants of
our simple example model.

In conclusion, the class of SUSY hybrid inflation models proposed
here not only solves the $\eta$-problem using a Heisenberg
symmetry, and stabilize the associated modulus during inflation,
but are also well motivated both from string theory and particle
physics, and leads to an acceptable value of the spectral index,
while predicting very small tensor modes. The specific example
of sneutrino hybrid inflation with no-scale SUGRA is
a particularly attractive possibility which deserves further study.

\section*{Acknowledgments}
S.F.K.\ acknowledges partial support from the following grants:
PPARC Rolling Grant PPA/G/S/2003/00096;
EU Network MRTN-CT-2004-50336;
EU ILIAS RII3-CT-2004-506222.
S.A.\ was partially supported by the the DFG cluster of excellence 
``Origin and Structure of the Universe''.
The work of
M.B.G.\ is done as part of the program {\sl Ram\'on y Cajal} of the  
Ministerio de Educaci\'on
y Ciencias (M.E.C.) of Spain. M.B.G.\ is also partially supported by
the M.E.C. under contract FIS
2007-63364 and by the Junta de Andaluc\'{\i}a group FQM 101.

\begin{appendix}
\section{Evolution of the Imaginary Parts of the Fields}
In section~\ref{Diagonal Basis}, we have used the assumption
that the evolution of the imaginary parts of the scalar components 
of all chiral superfields can be neglected.
Here, we show explicitly that this is justified for the phase
of $N$ and the imaginary part of the modulus $T$
from the full EOMs given that $s=h=0$. 
From Eq.~\eqref{KaehlermetricInflation}, we obtain the relevant
kinetic Lagrangian terms
\begin{equation}
\begin{split}
\mathcal{L}_{\text{Kin}}=&\left[f''(\rho)\,|N|^2-f'(\rho)\right]\,\del_{\mu}N\,\del^{\mu}N^*
-f'(\rho)\,N^*\,\del_{\mu}N\,\del^{\mu}T^*\\
&-f'(\rho)\,N\,\del_{\mu}N^*\,\del^{\mu}T + f''(\rho)\,\del_{\mu}T\,\del^{\mu}T^*\,.
\end{split}
\end{equation}
In the following we use the no-scale form~\eqref{specificf(rho)}
and decompose $T$ in its real and imaginary part.
Additionally, we write $N$ in terms of its modulus and phase
and introduce $\rho$ in terms of the real scalar DOFs:
\begin{equation}
T=\frac{1}{\sqrt{2}}\left(t_{\text{R}}+\text{i}\,t_{\text{I}}\right)\,,\quad
N=\frac{1}{\sqrt{2}}\,n\exp{\left(\text{i}\theta\right)}\,,\quad
\rho=\sqrt{2}\,t_{\text{R}}-\frac{1}{2}\,n^2\,.
\end{equation} 
Note the fact that using the definition of $\rho$, we
can fully eliminate $t_{\text{R}}$.
The full system is thus described by $(t_{\text{I}},\theta,\rho,n)$
with the kinetic terms given by
\begin{equation}
\mathcal{L}_{\text{Kin}}=\frac{3}{2\,\rho^2}
\left[\frac{{\dot{\rho}}^2}{2}+{\dot{t}}^2_{\text{I}}+\rho\,\dot{n}^2
+\frac{1}{2}\,n^4\,\dot{\theta}^2+\rho\,n^2\,\dot{\theta}^2
-\sqrt{2}\,n^2\,\dot{\theta}\,\dot{t}_{\text{I}}\right]\,.
\end{equation}
Since neither the tree-level nor the one-loop potential 
depend on $t_{\text{I}}$ and $\theta$, these are flat directions
and we have to make sure that they get ``frozen'' very quickly 
due to expansion and their EOMs decouple from the
$\rho\,$- and $n\,$-evolution. 
As the effective potential we apply Eq.~\eqref{effectivePot}
and obtain the coupled set of EOMs
\begin{equation}\label{fullEOMs}
\begin{split}
\ddot{t}_{\text{I}}+3\,H\,\dot{t}_{\text{I}}-2\,\frac{\dot{\rho}}{\rho}\,\dot{t}_{\text{I}}
-\frac{3}{\sqrt{2}}\,H\,n^2\,\dot{\theta}-\sqrt{2}\,n\,\dot{n}\,\dot{\theta}
-\frac{1}{\sqrt{2}}\,n^2\,\ddot{\theta}+\sqrt{2}\,\frac{\dot{\rho}}{\rho}\,n^2\,\dot{\theta}&=0\,,\\
\left(1+2\,\frac{\rho}{n^2}\right)\left[\ddot{\theta}+3\,H\,\dot{\theta}-2\,\frac{\dot{\rho}}{\rho}\,\dot{\theta}\right]+\\
\left[\left(4\,\frac{\dot{n}}{n}
+2\,\frac{\dot{\rho}}{n^2}+4\,\frac{\rho}{n^3}\,\dot{n}\right)\,\dot{\theta}
+\sqrt{2}\left(2\,\frac{\dot{\rho}}{\rho\,n^2}-2\,\frac{\dot{n}}{n^3}-\frac{3\,H}{n^2}\right)\,\dot{t}_{\text{I}}
-\sqrt{2}\,\frac{\ddot{t}_{\text{I}}}{n^2}\right]&=0\,,\\
\ddot{\rho}+3\,H\,\dot{\rho}-\frac{\dot{\rho}^2}{\rho}+\dot{n}^2+\frac{2\,\rho^2}{3}\,\frac{\del V_{\text{eff}}}{\del\rho}
+2\,\frac{\dot{t}^2_{\text{I}}}{\rho}
+n^2\,\dot{\theta}^2+\frac{n^4\,\dot{\theta}^2}{2\,\rho}-\frac{2\sqrt{2}}{\rho}\,n^2\,\dot{\theta}\,\dot{t}_{\text{I}}
&=0\,,\\
\ddot{n}+3\,H\,\dot{n}-\frac{\dot{\rho}}{\rho}\,\dot{n}+\frac{\rho}{3}\,\frac{\del V_{\text{eff}}}{\del n}
-n\,\dot{\theta}^2-\frac{n^3}{\rho}\,\dot{\theta}^2+\sqrt{2}\,\frac{n}{\rho}\,\dot{\theta}\,\dot{t}_{\text{I}}&=0\,.
\end{split}
\end{equation}
Note from the last two equations that in the 
limit $\dot{t}_{\text{I}}\rightarrow0$ and $\dot{\theta}\rightarrow 0$,
the evolution of $n$ and $\rho$ decouple from $t_{\text{I}}$ and 
$\theta$ and we recover Eqs.~\eqref{EOMs}.
\begin{figure}[!h]
\psfrag{ti}{$t_{\text{I}}$}
\psfrag{rh}{$\rho$} 
\psfrag{ni}{$n$}
\psfrag{im}{$\theta$} 
\psfrag{time}{\hspace{-0.7cm}$t\,[10^3\,\MP^{-1}]$}
\psfrag{MP}{\hspace{-1cm}$t_{\text{I}},\theta,\rho,n\,[\MP]$}
\center
\includegraphics[scale=0.9]{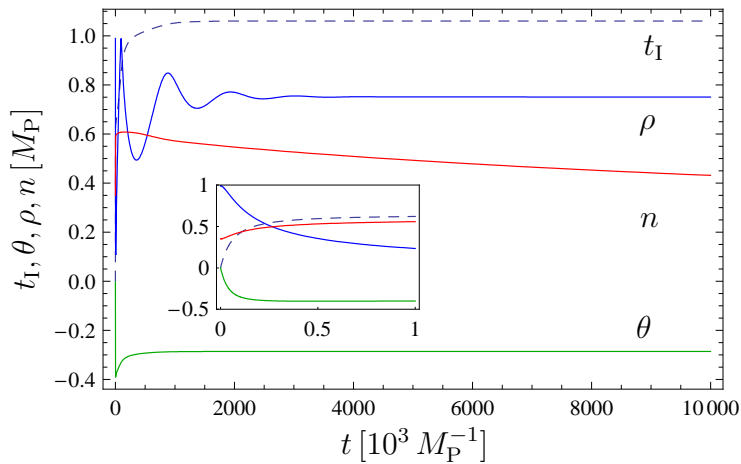}
\caption{\label{phases} Full evolution including the 
imaginary part $t_{\text{I}}$ and the phase $\theta$.
The purpose of the inlay is to show that for small $t$,
the evolution of the fields is perfectly smooth.}
\end{figure}

In the following, we simulate the full evolution 
described by Eq.~\eqref{fullEOMs} for some generic
initial conditions.
With the same renormalization scale $Q$ and
model parameters as in the simulation
in section~\ref{One-Loop Effective Potential},
the field evolution is plotted versus cosmic time
in Fig.~\ref{phases}.
As initial conditions for the fields, we chose
the values 
$\left(t_{\text{I}},\theta,\rho,n\right)|_{t=0}=
\left(0,0,0.99,0.25\right)$ and the velocities
$(\dot{t}_{\text{I}},\dot{\theta},\dot{\rho},\dot{n})|_{t=0}=
\left(0.01,-0.01,0,0\right)$.
As can be seen from the plot, both $t_{\text{I}}$
and $\theta$ obtain initial velocities in opposite directions. 
In this regime,
the evolution of $\rho$ and $n$ is influenced 
by them. However, due to the strong damping 
from the expansion of the Universe, after a 
very short period of time, both the imaginary 
part and the phase get frozen and stay
constant subsequently. Thereafter, the
$\rho$ and $n$ trajectories are not affected 
by $t_{\text{I}}$ and $\theta$ anymore.

We thus conclude that putting 
the phase of $N$ and $\text{Im}(T)$ to zero initially and using 
the decoupled Eqs.~\eqref{EOMs} for the absolute 
value and $\text{Re}(T)$ is justified. 
Similar conclusions have been drawn in ~\cite{Ellis:2006ara}.
\end{appendix}

\end{document}